\documentclass[showpacs,amsmath,amssymb,preprint]{revtex4-1} 
\pdfoutput=1
\usepackage{graphicx,anysize}
\usepackage{amsmath}
\usepackage{amssymb}
\usepackage{psfrag}
\usepackage{threeparttable}
\usepackage{float}
\usepackage{color}
\usepackage{varioref}
\usepackage{epic}
\usepackage{eepic}
\usepackage{makeidx}
\usepackage{epsfig}
\usepackage{supertabular}
\usepackage{hyperref}
\usepackage{amsmath}
\usepackage{multirow}
\usepackage{booktabs}
\hypersetup{colorlinks,citecolor=blue,linkcolor=red,urlcolor=blue}
\def \kbt{$k_{\rm B}T$}
\begin{document}

\title{Application of a Free Energy Landscape Approach to Study Tension Dependent Bilayer Tubulation Mediated by Curvature Inducing Proteins}

\author{Richard W. Tourdot}
 \email{tourdot@seas.upenn.edu}
\affiliation{Department of Chemical and Biomolecular Engineering, University of Pennsylvania, Philadelphia, PA, 19104}
\author{N. Ramakrishnan}
 \email{ramn@seas.upenn.edu}
\affiliation{Department of Bioengineering, University of Pennsylvania, Philadelphia, PA, 19104}
\author{Tobias Baumgart}
\email{baumgart@sas.upenn.edu}
\affiliation{Department of Chemistry, University of Pennsylvania, Philadelphia, PA, 19104}
\author{Ravi Radhakrishnan}
 \email{rradhak@seas.upenn.edu}
\affiliation{Department of Chemical and Biomolecular Engineering, Department of Bioengineering, University of Pennsylvania, Philadelphia, PA, 19104}

\date{\today}
\begin{abstract}
We investigate the phenomenon of protein induced tubulation of lipid bilayer membranes within a continuum framework using Monte Carlo simulations coupled with the Widom insertion technique to compute excess chemical potentials. Tubular morphologies are spontaneously formed when the density and the curvature-field strength of the membrane bound proteins exceed their respective thresholds and this transition is marked by a sharp drop in the excess chemical potential. We find that the planar to tubular transition can  be described by a micellar model and that the  corresponding free energy barrier increases with increase in the curvature-field strength, (i.e. of protein-membrane interactions), and also with increase in membrane tension.
\end{abstract}

\preprint{To appear in Phys.  Rev. E}

\maketitle
\section{Introduction}
Highly curved membrane structures at the tens-of-nanometers length scale, such as buds, vesicles, and tubules are 
essential functional intermediates in cell physiological processes.  These 
intermediates are orchestrated by the membrane remodeling activities of a specialized class of proteins 
\cite{McMahon:2005km,Kozlov:2014jv,Shibata:2009je,Zimmerberg:2005jka,Farsad:2003cm,Ayton:2009gw,Chen:1998jga,Farsad:2001jj}.  Proteins comprised of Bin-Amphiphysin-Rvs (BAR), epsin-N-terminal homology (ENTH), and inverted-BAR (IBAR) domains are enriched in cellular pathways involving traffic and transport in cells \cite{Boucrot:2015ie,McMahon:2005km}. It is shown that these protein-domains induce membrane curvature on a lipid membrane bilayer \cite{Zimmerberg:2005jk,McMahon:2005km}; when multiple proteins are localized to a region, they act cooperatively to induce/ stabilize the afore mentioned morphologies that are otherwise unstable. Disc-like shapes in the endoplasmic reticulum has been shown to be stabilized by DP1 (deleted-in-polyposis) and reticulon class proteins \cite{Shibata:2010fz}, while membrane tubules are induced through ENTH domains \cite{Ford:2002if}, BAR domains \cite{Zimmerberg:2005jk,McMahon:2005km}, dynamin \cite{Hinshaw:2000fi}, Shiga toxin \cite{TRA:TRA1116}, and other proteins such as Exo70 \cite{Zhao:2013hi}.

The molecular interaction of a curvature inducing protein with a bilayer membrane has been extensively studied using all atom and coarse grained simulations for a various classes of curvature remodeling proteins. These studies can be broadly classified into those that focus on the properties of the curvature field at the molecular scale~\cite{Blood:2006uc,Yin:2009jz,Zhao:2013hi,Lai:2012hk} and those at focus on their membrane remodeling effects at the mesoscale~\cite{Reynwar:2007kma,Cui:2011ei,Cui:2013dx,Lyman:2010jm,Arkhipov:2008fg}. On the other hand, at the continuum scale, elasticity based theoretical and computational models have been used to study membrane remodeling by treating the individual proteins as an inclusion that modulates the curvature of the membrane surface~\cite{Kim:1998ft,Sens:2008dz,Brown:2011iq,Bahrami:2012gb,Saric:2012hb,Dasgupta:2013iz,Lipowsky:2012gv,Rangamani2014751,Ramakrishnan20141}. Conventionally, the elastic Hamiltonian (see eqn.\eqref{eqn:Helf-continuum}) governing the energy of the membrane is taken to be the free energy of the system and in cases where membrane inclusions are also considered the conformational entropy of these inclusions are accounted for by treating them as interacting particles with well defined mixing energies~\cite{Leibler:1986kq,Auth:2009ksa,Baumgart:2011en,Singh:2012gb,Sorre:2012if,Aimon:2014if}. However, in the context of thermodynamics, the true free energy should also account for the entropic contributions from the membrane degrees of freedom, which would involve explicit free energy calculations that also account for thermal fluctuations of the system~\cite{Frenkel2001}. For example, an umbrella sampling based coarse grained molecular simulation has been used to determine the polymerization free energy of BAR domain protein on membranes with varying tension~\cite{Simunovic:2015bx}. Recently we have introduced a number of free energy methods derived from chemical physics \cite{Frenkel:2001} to delineate the free energy landscapes of membranes remodeled by curvature inducing proteins~\cite{Ramakrishnan20141,Tourdot:2014hm,Tourdot:2014ef}. In this article we use some of these methods to predict the stability of emergent morphologies such as tubules, blebs, and buds that arise due to the cooperative interactions of the proteins with the membrane.



Two theories based on stability/instability have been advocated to address the role of cooperativity. Leibler and others~\cite{Leibler:1986kq,Sens:2004ee,Shi:2014dx} have proposed that the presence of these proteins 
generates a curvature instability, which drives a morphological transition in the liposome, the onset of which is related directly to the strength of the induced-curvature field. The authors have 
developed an analytical model to describe the boundary that separates the planar and tubular  regions; the boundary depends on factors such as membrane bending rigidity, tension, and induced-field strength. Sorre \textit{et.~al.} \cite{Sorre:2012if} presented a thermodynamic theory (accounting for the protein's translational entropy on the membrane surface) that quantifies the force acting on a tether pulled from a giant unilamellar vesicle in the presence of a curvature-coupling protein. However, the theory idealizes the emergent membrane geometry to be that of a cylinder attached to a flat membrane.

Alternatively, tour-de-force coarse-grained molecular dynamics calculations of membranes decorated with oligomerized networks of ENTH~\cite{Lai:2012hk}, 
N-BAR~\cite{Yin:2009jz}, and Exo70~\cite{Zhao:2013hi} domains have shown that in the presence of these 
proteins tubular and vesicular morphologies are stable. A similar approach has been used to investigate the effect of 
protein aggregation, cooperative interactions, and membrane elasticity~\cite{Simunovic:2013ca,Simunovic:2015bx} on the 
formation of highly curved membrane morphologies. The first class of models 
utilize a continuum top-down approach to determine regions of curvature instability and have limited 
capabilities in predicting emergent morphologies. The second class of models utilize a bottom-up molecular approach to study microscopic mechanisms governing protein 
oligomerization and membrane remodeling,  but do not directly compare the thermodynamic stabilities of the planar and tubular states. 


Open questions relevant to cell physiology still remain unanswered and include: what is the nature of the emergent morphological state (cylinder, bud, bleb etc.), and what are the morphological features at the mesoscale (e.g., protein 
density and organization, geometry)? What is the thermodynamic free energy landscape defining these morphological states and their relative stabilities, the driving forces governing these transitions (e.g., energetic vs. entropic costs of driving membrane curvature)? More significantly, what are the roles of direct and membrane-mediated cooperative interactions of proteins in defining the transition free energy landscapes, (e.g., curvature contribution to the chemical potential determines protein recruitment by which curvature gradients define the driving force for transport). 

Recent experimental work by Shi and Baumgart~\cite{Shi:2014ho} have brought the focus back to these questions, where they report a reversible transition between the tubule and planar states, which is  strongly influenced by protein surface density and membrane tension. It is becoming clear that the precise control of spatial localization and temporal dynamics of the curvature-inducing proteins is crucial not only to the regulation of membrane-mediated trafficking such as endocytosis~\cite{Tourdot:2014hm}, exocytosis~\cite{Zhao:2013hi}, but also in cell migration ~\cite{Tsujita:2015kr}. The physical microenvironment around a cell such as matrix stiffness and dimensionality will influence the physical variables on the membrane such as membrane stiffness or tension \cite{DizMunoz:2013bi}, and will dictate the underlying trafficking and migratory stimuli in such cells mediated by curvature inducing proteins. 

 
\section{Methods}
We address the biophysical challenges discussed above by utilizing  a mesoscale computational model we have developed to describe protein-induced tubulation and combining it with methods to delineate the free energy landscapes of protein-recruitment and membrane morphological transitions~\cite{Tourdot:2014ef,Ramakrishnan20141,Tourdot:2014hm}.  The core methodology for performing the simulations and free energy calculations are essentially the same as that reported in~\cite{Tourdot:2014ef}.  Here, we recapitulate only the essential details and enhancements to the methodology.
\subsection{Continuum model for membrane and protein induced spontaneous curvature field} Following the approaches in our previous works \cite{Ramakrishnan20141,Tourdot:2014hm,Tourdot:2014ef}, the 
membrane is modeled as a thin elastic sheet, which is discretized into a triangulated mesh with $N$ vertices and $T$ 
triangles \cite{Ramakrishnan:2010hk}. The energy of this surface is given by the discretized form of the Canham-Helfrich 
Hamiltonian \cite{Helfrich:1973td}, 
\begin{equation}
\mathcal{H} = \sum_{v=1}^{N} \left\{\frac{\kappa}{2} \left({C}_{1,v}+{C}_{2,v} - H_{0,v}\right)^2  + \sigma_{\rm bare} 
\right \} A_v.
\label{eqn:Helf-continuum}
\end{equation}
$\kappa$ and $\sigma_{\rm bare}$ are the bending rigidity and bare surface tension of the 
membrane~\cite{Tourdot:2014ef,Ramakrishnan20141}. ${C}_{1,v}$ and ${C}_{2,v}$ are the principal curvatures at vertex 
$v$, computed as in \cite{Ramakrishnan:2010hk}, while $A_v$  denotes the corresponding surface area. Protein induced 
curvature remodeling effects are included through the spontaneous curvature field $H_{0,v}$. If ${\bf r}_v$ denotes the 
position of vertex $v$ and ${\bf R}_{i}$ denotes the position of protein $i$, then the effective spontaneous curvature 
at $v$, due to the $n_P$ proteins on the surface, is  computed as:
\begin{equation}
H_{0,v}=\sum_{i=1}^{n_P} C_0 \exp\left(-{({\bf 
r}_v-{\bf R}_i)^2}/{2\epsilon^2}\right).
\label{eqn:sponcurv}
\end{equation}
Both the membrane and protein degrees of freedom evolve through the coupled set of dynamically triangulated Monte Carlo moves described in~\cite{Tourdot:2014ef}. There is no explicit interaction between protein fields besides a self avoidance potential that prevents two protein fields being localized to the same  vertex of the triangulated surface.  All other protein interactions are mediated through the Helfrich Hamiltonian. The results presented here are for a membrane surface with $N=900$ vertices, $\kappa=20k_BT$ 
and $\sigma_{\rm bare}=0$. In our previous work \cite{Tourdot:2014ef}, we had noted that this model predicts a 
tubulation transition. In the following discussion, we present our analysis of the tubulation transitions as  a function 
of the magnitude of the spontaneous curvature ($C_0$), its variance ($\epsilon^2$), the number of proteins on the 
membrane ($n_P$) and  the excess area of the membrane ($A/A_p$) \textemdash defined as the ratio of the curvilinear area 
($A$) to its projection onto the $x-y$ plane ($A_p$).  All curvatures are presented in units of $a_0^{-1}$ with $a_0 =10 
{\rm nm}$. The choice of the model parameters including their method of estimation and justification is based on experimental data, and the computational details regarding the simulations are available in our previous work~\cite{Tourdot:2014ef}. 

\subsection{Inhomogeneous Widom insertion} \label{sec:inhomogeneous}
The behavior of the remodeled membrane is quantified in terms of the excess chemical potential $\mu^{ex}$ for $n_P$ protein-fields and is computed using the Widom field insertion technique \cite{Tourdot:2014ef} as 

\begin{equation}
\mu^{ex} = -k_B T \ln{\int{\left \langle e^{-\beta \Delta {\cal H}} \right \rangle_{M} {\cal P}(s_{M+1}) ds_{M+1}}}.
\label{eqn:muex}
\end{equation}
Here $\Delta {\cal H} = {\cal H}\left({M+1}\right) - {\cal H}\left({M}\right)$ where $M$ denotes the number of proteins on the membrane, $s_{M}$ denotes the corresponding conformational space of the system and ${\cal P}$ is the probability density to add the $M+1^{\rm th}$ protein field at site $s_{M+1}$ which is taken to be uniform. The excess chemical potential in eqn.~\eqref{eqn:muex} is an average value which corresponds to the chemical potential measured in bulk, while the same formulation can also be extended to systems with spatially varying density~\cite{Frenkel:2001}. In this article, we extend the simulation methodology from \cite{Tourdot:2014ef} to compute spatially dependent excess chemical potentials.  If ${\bf r}$ denotes a state point in the configurational phase space, $\mu^{ex}({\bf r})$ its chemical potential, and $\Delta {\cal H}({\bf r})$ the energy change at ${\bf r}$ due to the insertion of the $(M+1)^{\rm th}$ protein at any point on the membrane, then the spatially varying excess chemical is given by 
\begin{equation}
\mu^{ex}({\bf r}) = -k_B T\ln \int \left \langle {e}^{-\beta\Delta {\cal H} ({\bf r})} \right \rangle_{M} {\cal P}(s_{M+1}) ds_{M+1}.
 \label{eqn:inhomo}
\end{equation}
In this study, ${\bf r}$ is binned (histogrammed) based on the values of the mean curvature at different spatial locations, $H_{v} = (C_{1,v} + C_{2,v})/2$, at each vertex $v$ where the test-protein-field is inserted. The tubular regions on the  membrane are identified based on the bimodal distribution in the histograms of mean curvature, as described in Sec.~\ref{sec:results}.  
In order to achieve adequate sampling for inhomogeneous Widom insertion calculations, each membrane simulations are run for at least 3 million Monte Carlo steps. Data for Widom  test-field-insertion is collected only during the production phase which corresponds to the second half of the simulation (i.e. the last 1.5 million MC steps) in order to ensure membrane equilibration.  In specific, the test-protein-field is inserted every 100 MC steps at randomly chosen spatial locations (here we have limited the maximum number of locations to 20) with the value of $\exp(-\beta \Delta {\cal H}(r))$ being recorded for every insertion move.  The reported values of the error bars in $\mu^{ex}$ correspond to the standard deviation  computed over four replicate ensembles.

\subsection{Computing membrane tension from the undulation spectrum}\label{sec:uspec}
A planar membrane is characterized by the extensive variables entropy ($S$), surface area ($A$), projected area ($A_{P}$), and the number of protein fields ($n_P$). 
If $\gamma$ is the tension due to the frame (also called the frame tension), $\mu_{m}$ is the chemical potential of the membrane, $\mu$ is the chemical potential of the protein field, and $T$ is the temperature, then at constant projected area ($A_p$) the suitable thermodynamic potential is given by,
\begin{equation}
\begin{split}
dF(N,n_P,\sigma,A_p,T)= \mu_{m} dN + \mu dn_P - A d \sigma +\gamma dA_{p}  - SdT.
\end{split}
\label{eqn:pl-free-ener}
\end{equation} 
In this ensemble we initialize the system with set values of $N$, $n_P$, $A_P$,  and $T$.  The surface tension $\sigma$ represents the renormalized tension which can be estimated through the fluctuation spectrum analysis discussed below.


The membrane is initialized in a 30 by 30 hexagonal lattice with a link length, $l$, which can vary within the range of self avoidance constraints $a_0 $ and $\sqrt{3} a_0$.  The initial link length sets the membrane projected area according to $A_p = 900 (l a_0)^2 \sqrt{3}/2$. Upon equilibration, thermal undulations tend to increase the curvilinear area of the membrane (i.e. $A \geq A_p$) and this defines an excess area reservoir which is dependent on the value of $l$. Hence, the entropic tension depends on the value of the excess area reservoir, $A/A_{p}$, which can be measured by analyzing the power spectrum of membrane undulations \cite{Tourdot:2014ef}.
In the absence of any spontaneous curvature field the power spectrum is given by,
\begin{equation}
k_{B}T = \langle h_{q}h_{-q} \rangle A_{p}\left[\kappa q^{4}+\sigma q^{2} \right ].
\label{eqn:simple-hqhq}
\end{equation}
Eqn.~\eqref{eqn:simple-hqhq} can be used to measure the renormalization behavior of $\kappa$ and $\sigma$ as a function of $A/A_{p}$ as discussed in \cite{Tourdot:2014ef}. However, this simple relationship does not hold for a membrane with $n_{P} >0$. In such a scenario the contributions from the spontaneous curvature fields to the power spectrum should also be accounted for. The power spectrum which incorporates the effect of the protein spontaneous curvature fields has been previously derived in Ref.~\cite{Tourdot:2014ef} and is given by,
\begin{multline}
\langle {\cal H} \rangle = \frac{A_p}{2} \, \sum_{\vec{q}} \sum_{\vec{q'}} \, \{  [  q^2 {q'}^2   \langle h_{q} h_{q'} \rangle -  q^2 \langle h_{q} h_{0,q'} \rangle \\
-  {q'}^2 \langle h_{0,q} h_{q'} \rangle +  \langle h_{0,q} h_{0,q'} \rangle  ]  \kappa_{q+q'} \, + \, q q' \, [ \langle h_{q} h_{q'} \rangle  ] \, \sigma_{q+q'} \}.
 \label{eqn:hqhmq-complex}
\end{multline}
Here $q$ and $q'$ correspond to two independent modes which are coupled to each other through the elastic parameters $\kappa_{q+q'}$ and $\sigma_{q+q'}$ which represent the mode specific bending rigidity and tension. $h_{0,q}$ is the Fourier transform of the spontaneous curvature field $H_0 ({\bf r})$.  While this formalism for carrying out the fluctuation spectrum analysis in the presence of a finite number of non-zero curvature fields was presented in ~\cite{Tourdot:2014ef}, its practical utility was not demonstrated. Here we apply this formalism and show that it can be utilized to compute the renormalized values of $\kappa$ and $\sigma$ in the presence of spontaneous curvature.  For a homogeneous distribution of $\kappa$ and $\sigma$, $\kappa_{q+q'}=\kappa \delta_{{q,q'}}$ and $\sigma_{q+q'}=\sigma \delta_{{q,q'}}$ and eqn.~\eqref{eqn:hqhmq-complex} reduces to
\begin{multline}
\langle {\cal H} \rangle = \frac{A_p}{2} \, \sum_{\vec{q}} \, \{  [  q^4   \langle h_{q}^2 \rangle -  q^2 \langle h_{q} h_{0,q} \rangle -
{q}^2 \langle h_{0,q} h_{q} \rangle +  \langle h_{0,q}^2 \rangle  ]  \kappa \, + \, q^2 \,  \langle h_{q}^2 \rangle \, \sigma \}. 
 \label{eqn:hqhmq-complex-2}
\end{multline}
Each of the modes obey equipartition and hence the relation for the power spectrum in terms of the various Fourier modes is given by
\begin{multline}
k_B T= A_p \{  [  q^4   \langle h_{q}^2 \rangle -  q^2 \langle h_{q} h_{0,q} \rangle  - 
{q}^2 \langle h_{0,q} h_{q} \rangle +  \langle h_{0,q}^2 \rangle  ]  \kappa \, + \, q^2 \,  \langle h_{q}^2 \rangle \, \sigma \}.
\label{eqn:hqhmq-complex-3}
\end{multline}
The renormalized values of $\kappa$ and $\sigma$, in the presence of spontaneous curvature inducing protein fields, can be determined through a nonlinear fit of eqn.\eqref{eqn:hqhmq-complex-3}.  

\section{Results and discussion}\label{sec:results}
\subsection{Tubulation and bimodal distribution of  membrane mean  curvature}
\begin{figure*}[!ht]
\includegraphics[width=15cm]{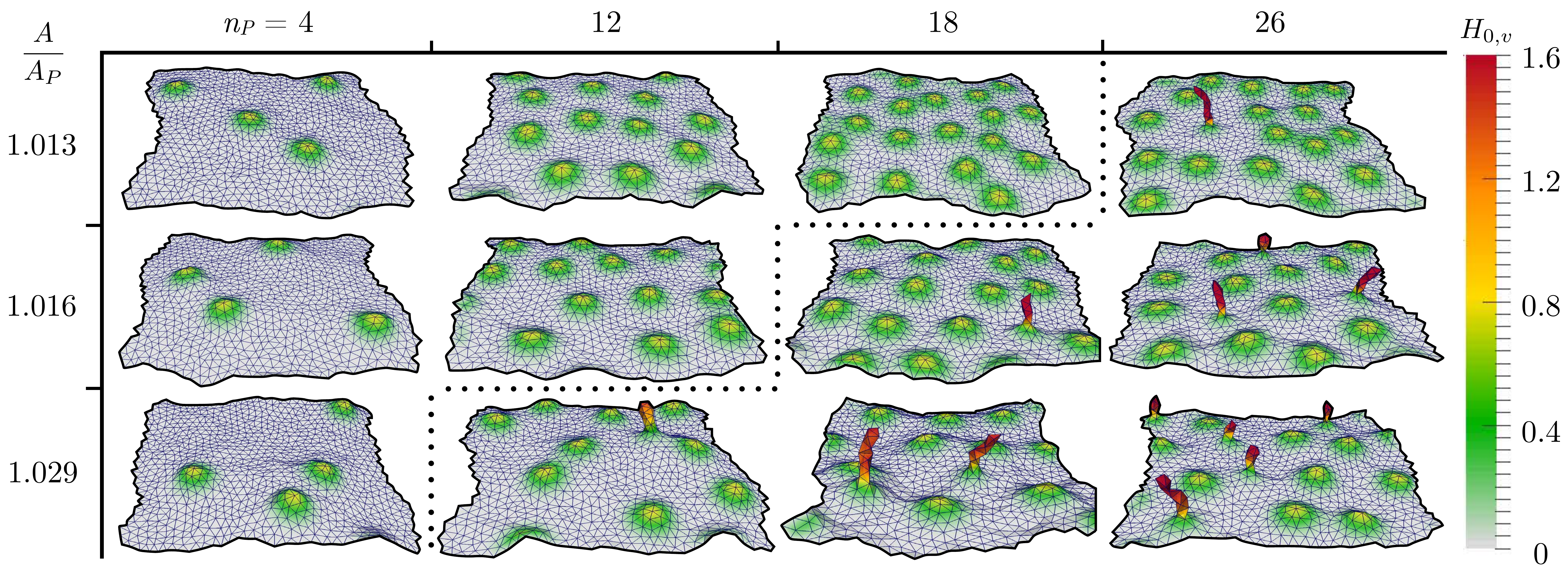}
\caption{(color online). Representative snapshots of equilibrium membrane morphologies as a function of $n_P$ and $A/A_p$. The membrane surfaces are colored based on the value of $H_{0,v}$  (expressed in units of $a_0^{-1}$) \textemdash an isolated Gaussian bump represents an individual protein field while tubules, formed by the aggregation of multiple protein fields, are seen as sharp protrusions. All protein fields shown have the parameters $C_0 = 0.8 \, a_0^{-1} $ and $\epsilon^2 = 6.3 \, a_0^{2}$. \label{fig:collage}} 
\end{figure*}

A membrane surface can display a number of equilibrium shapes that depend on the bending stiffness, excess area, and the number of curvature inducing proteins on its surface. Snapshots of the various conformations of a membrane with $\kappa=20$ \kbt{} as a function of $A/A_p$ and $n_{P}$ are shown in Fig.~\ref{fig:collage}. It can be seen that the equilibrium shapes vary between smooth planar conformations, for small $A/A_p$ or $n_P$, and rough protrusions for large $A/A_p$ or $n_P$. 

\begin{figure}[!h]
\includegraphics[width=7.5cm]{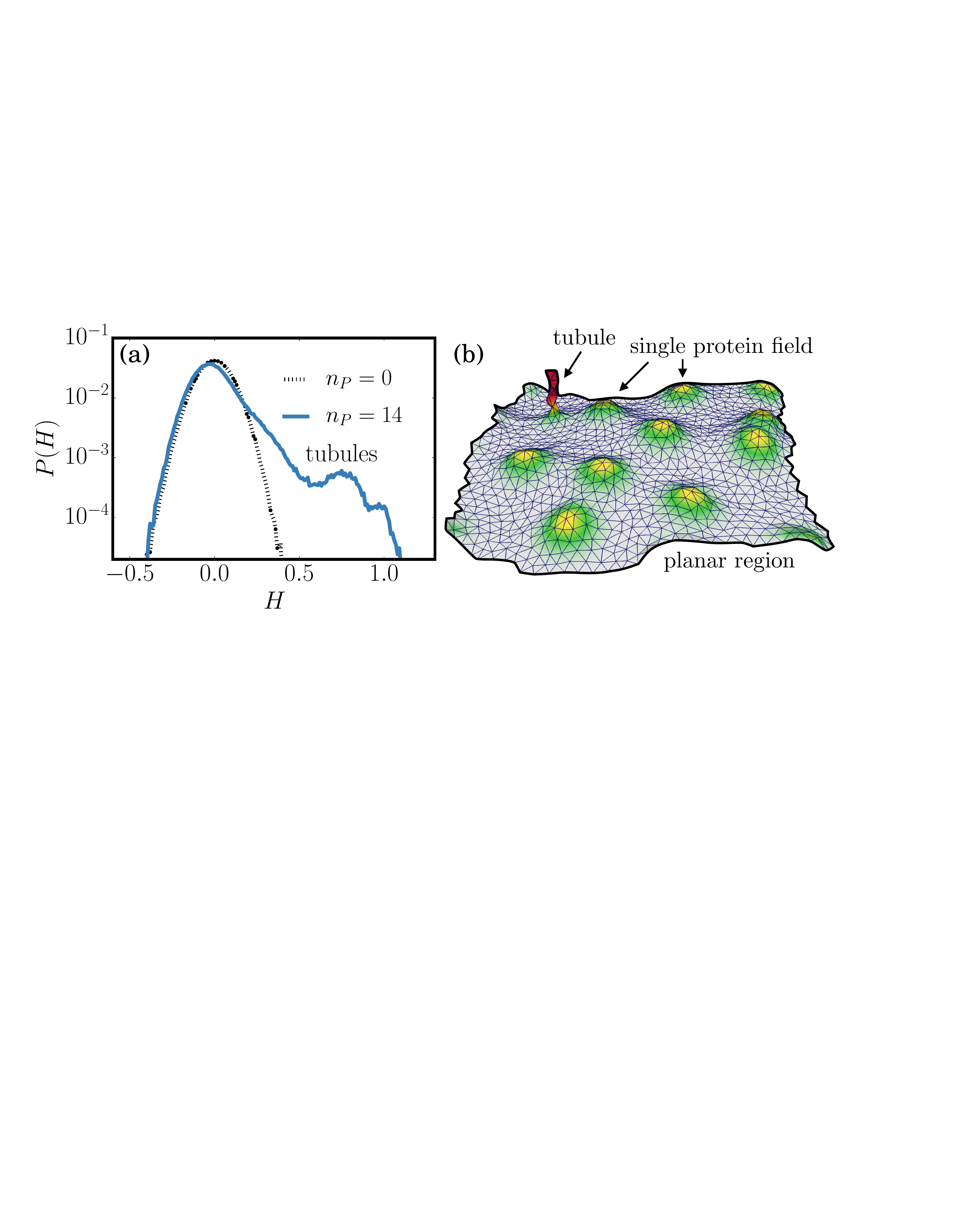}
\caption{(color online) a) Probability density of the membrane mean curvature for two protein concentrations, $n_P = 0$ and $14$, for a protein field with $C_0=0.8$ and $\epsilon^2=6.3$. b) Snapshot corresponding to the membrane with $n_P=14$, that clearly illustrates co-existing planar and tubular regions on the membrane. \label{fig:tube-histogram}}  
\end{figure}

\begin{figure}[!h] 
\centering
\includegraphics[width=7.5cm,clip]{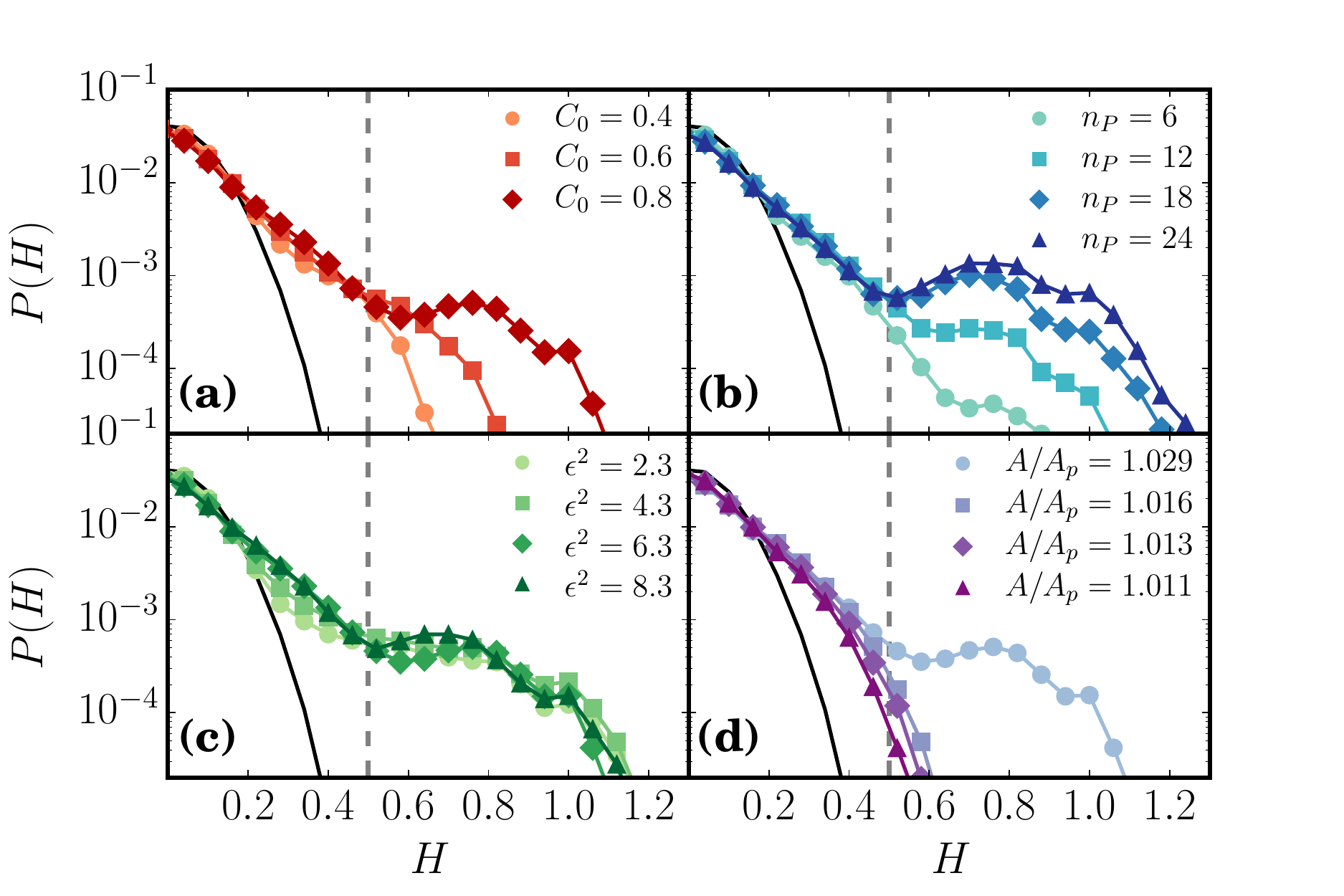}
\caption{(color online). Histograms of mean curvature for simulations with: (a) a range of peak spontaneous curvatures $C_0$, (b)  several protein concentrations $n_P$, (c) a range of curvature field extents $\epsilon^2$, and (d) several different membrane excess areas $A/A_p$.  All panels have the parameters $C_0 = 0.8 a_0^{-1}$, $\epsilon^2 = 6.3 a_0^2$,  $n_P = 14$, and $A/A_p = 1.029$ unless otherwise stated.  Mean curvature cutoff of $0.5 a_0^{-1}$ shown as vertical dotted line.
\label{fig:mc-histograms}}  
\end{figure}

In our simulations, a tubule is a protrusion above the mean surface of the membrane, as observed in  Fig.~\ref{fig:collage}. The tubulation transition itself is marked by the onset of a bimodal distribution of the mean curvature, $P(H)$, as depicted in Fig.~\ref{fig:tube-histogram} for $\kappa=20$ \kbt{}, $A/A_p=1.029$, for two protein concentrations $n_P=0$ and $14$ with $C_0=0.8a_0^{-1}$. The characteristic peaks at  $H=0$ and $H>0.5$ seen for $n_P=14$ correspond to planar and tubular regions, respectively, and the peak at higher mean curvatures is not observed for dilute protein concentrations (data shown for $n_P=0$). Furthermore, Figs.~\ref{fig:mc-histograms}(a-d) show the distribution of mean curvature as a function of $C_0$, $n_P$, $\epsilon^2$, and $A/A_p$ respectively.     It is evident that the tubulation transition is a function of the various parameters that characterize the membrane-protein system. In Fig.~\ref{fig:mc-histograms}, the absence of a bimodal distribution indicates that the curvature remodeling effects are not strong enough to 
stabilize tubular structures, and collectively the results indicate that the tubulation transition occurs only above a threshold protein concentration, which is strongly influenced by both the characteristics of the protein field \textemdash~ given by $C_0,\,\epsilon^{2}$ \textemdash~ and by the excess membrane area, $A/A_p$.
The curvature distribution $P(H)$ is a useful marker of tubulation, but can only be used unambiguously when a 
large number of tubules are present. Also, its ability to predict the tubulation boundary is limited when non-tubular structures such as blebs, buds, etc. are present. This is evident from examining the $P(H)$ versus  $n_P$, as shown in Fig.~\ref{fig:mc-histograms}(b); though $P(H)$ shows a clear bimodal distribution only above $n_P = 12$, the protrusions appear even for $n_P=10$, but the mode at larger values of $H$ does not appear since these structures are not persistent.  Hence, to faithfully resolve the transition boundary, we have computed the excess chemical potential, in order to quantify the nature of membrane tubule formation induced by curvature remodeling proteins.

\subsection{Excess chemical potentials as markers of tubulation}
In particular, we utilize the inhomogeneous Widom insertion technique (described in Sec.~\ref{sec:inhomogeneous}), which for our purpose involves the computation of three different excess chemical potentials, namely: (a) $\mu^{ex}$ in the entire system,  (b) $\mu_p^{ex}$ in spatial regions where $H<0.5$, and (c) $\mu_t^{ex}$ corresponding to the tubular regions, i.e. for regions with $H\geq 0.5$. The thresholds are consistent with (and derived from) the 
cutoff value ($H=0.5$) that separates the two modes in the $P(H)$ distributions, see Fig.~\ref{fig:mc-histograms}.

\begin{figure}[h] 
\centering
\includegraphics[width=7.5cm]{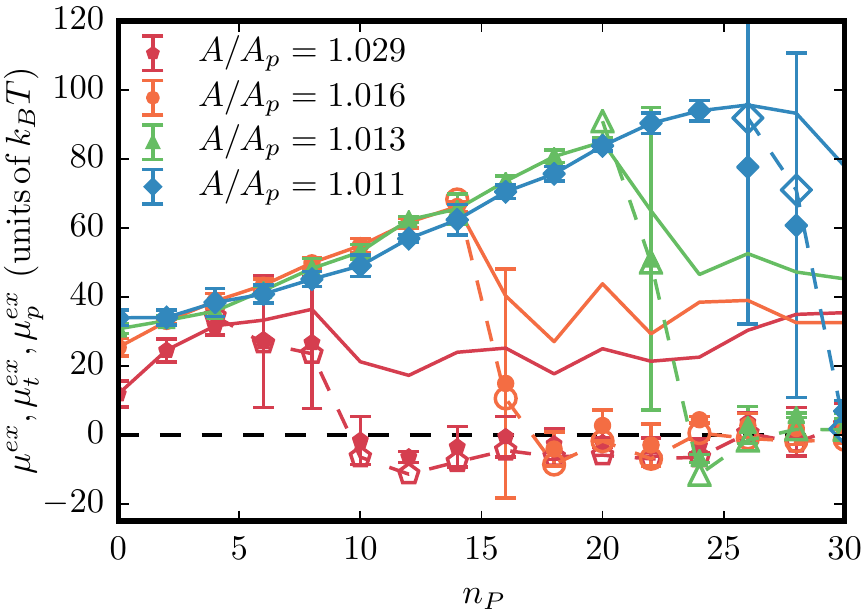}
\caption{(color online). The various excess chemical potentials as a function of $n_P$, for four values of $A/A_{p}$. 
For each value of $A/A_p$, filled symbols with error bars denote $\mu^{ex}$, open symbols with dotted lines represent 
$\mu_t^{ex}$, and solid lines correspond to $\mu_p^{ex}$. \label{fig:mu-excess}}  
\end{figure}

\begin{figure}[!h] 
\centering
\includegraphics[width=7.5cm,clip]{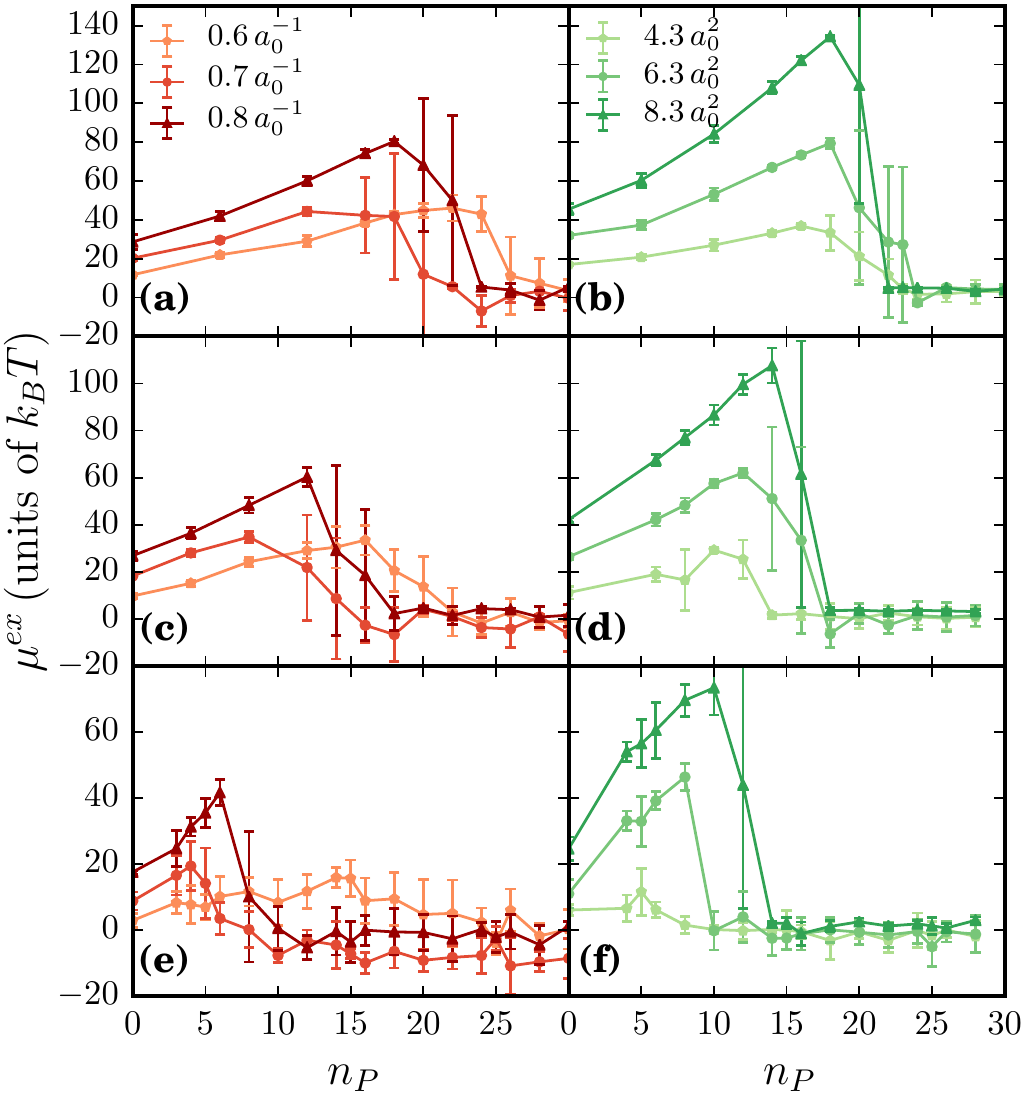}
\caption{(color online). Plot of the excess chemical potential vs protein number for a range of both $C_0$ and $\epsilon^2$ for several initial excess areas.  Solid lines with correspond to $\mu_p^{ex}$ while points with error bars correspond to $\mu^{ex}$. Panels a,c, and e depict data for a range $C_0$ with $\epsilon^2 = 6.3 \,a_0^2$ and corresponding excess areas (a) $A/A_p = 1.013$, (c) $A/A_p = 1.016$, and (e) $A/A_p = 1.029$.  Panels b,d, and f depict data for a range $\epsilon^2$ with $C_0 = 0.8 \,a_0^{-1}$ and corresponding excess areas (b) $A/A_p = 1.013$, (d) $A/A_p = 1.016$, and (f) $A/A_p = 1.029$. The values of $\mu_t^{ex}$ are similar to that of $\mu^{ex}$ and hence are not shown for clarity. \label{fig:mu-excess-parameters-bifurcation}}  
\end{figure}

The equilibrium chemical potential $\mu^{ex}$ as a function of $n_P$, for protein induced curvature field-strength of $C_0=0.8a_0^{-1}$ and $\epsilon^2=6.3a_0^2$, for different values of the membrane excess area is shown in Fig.~\ref{fig:mu-excess}. Shown alongside are the corresponding values of the excess chemical potentials: planar region $\mu_p^{ex}$ vs. tubular region $\mu^{ex}_t$. We note that in an inhomogeneous phase showing  spatial variation of density, the total chemical potential $\mu$ is a constant, which is the sum of  $\mu^{ex}$, which strongly depends on the underlying curvature at a given location and $\mu^{id}(\rho)$, which depends on the density at the location. When $n_P < 5$ the total excess chemical potential $\mu^{ex}$ is indistinguishable from the chemical potential obtained from the planar region $\mu_p^{ex}$, as is clearly seen for the case of $A/A_p = 1.029$.  However, at the onset of tubulation where $\mu_t^{ex}$ is well defined, $\mu^{ex}$ is slaved to the values of $\mu_t^{ex}$. This relation holds for all parameter values that can induce membrane tubules, and this is shown for a range of $C_0$, $\epsilon^2$, and $A/A_p$ in Fig.~\ref{fig:mu-excess-parameters-bifurcation}. 

The similarity in the values of $\mu^{ex}$ (the excess chemical potential in bulk) and $\mu_t^{ex}$ (the excess chemical potential in the tubular region) indicates the presence of a strong thermodynamic driving force to form tubulated regions on the membrane.  The transition behavior shows a bifurcation in the excess chemical potential versus density plane, and the transition point for a given field-strength of curvature induction is a function of the membrane excess area, $A/A_p$. As $n_P$ increases in the build-up to the transition $\mu^{ex}$ increases owing to repulsion between the protein fields. However, beyond the transition point $\mu^{ex}$, $\mu_p^{ex}$, and $\mu_t^{ex}$  decrease. The observed decrease in $\mu_t^{ex}$ in the tubular phase reflects that fact that the curvature contribution to $\mu^{ex}$ from the large mean curvatures of the tubule dominates the free energy contribution. That the $\mu_p^{ex}$ for the planar phase also drops (albeit by a much smaller amount relative to its value prior to the transition) is a reflection of the fact that the average density of the protein-fields in the planar region is a constant and lower than the protein density just prior to the transition. This observation can be rationalized by the fact that post-transition, addition of new protein fields results in their incorporation in the tubular phase keeping the density in the planar phase at a constant value, (see Fig.~\ref{fig:mu-excess}). That the fluctuations in the $\mu^{ex}$ values are higher at the transition region and are considerably lower pre- and post- transition along the $n_P$ axis has to do with sampling rather than any onset of criticality. This is reconciled through the $P(H)$ distributions which show metastability in the free energy landscape of the planar versus tubule phases, which is a not feature of a first-order-like transition. Moreover, as we discuss below, the transition we observe in the model is a state transition (akin to a micellar transition), and several features in our results outlined in Fig.~\ref{fig:mu-excess} are in striking agreement with analogous behavior reported for micellar systems.

\subsection{Membrane tubulation and its analogy to micellization}
\begin{figure}[!h] 
\centering
\includegraphics[width=7.5cm,clip]{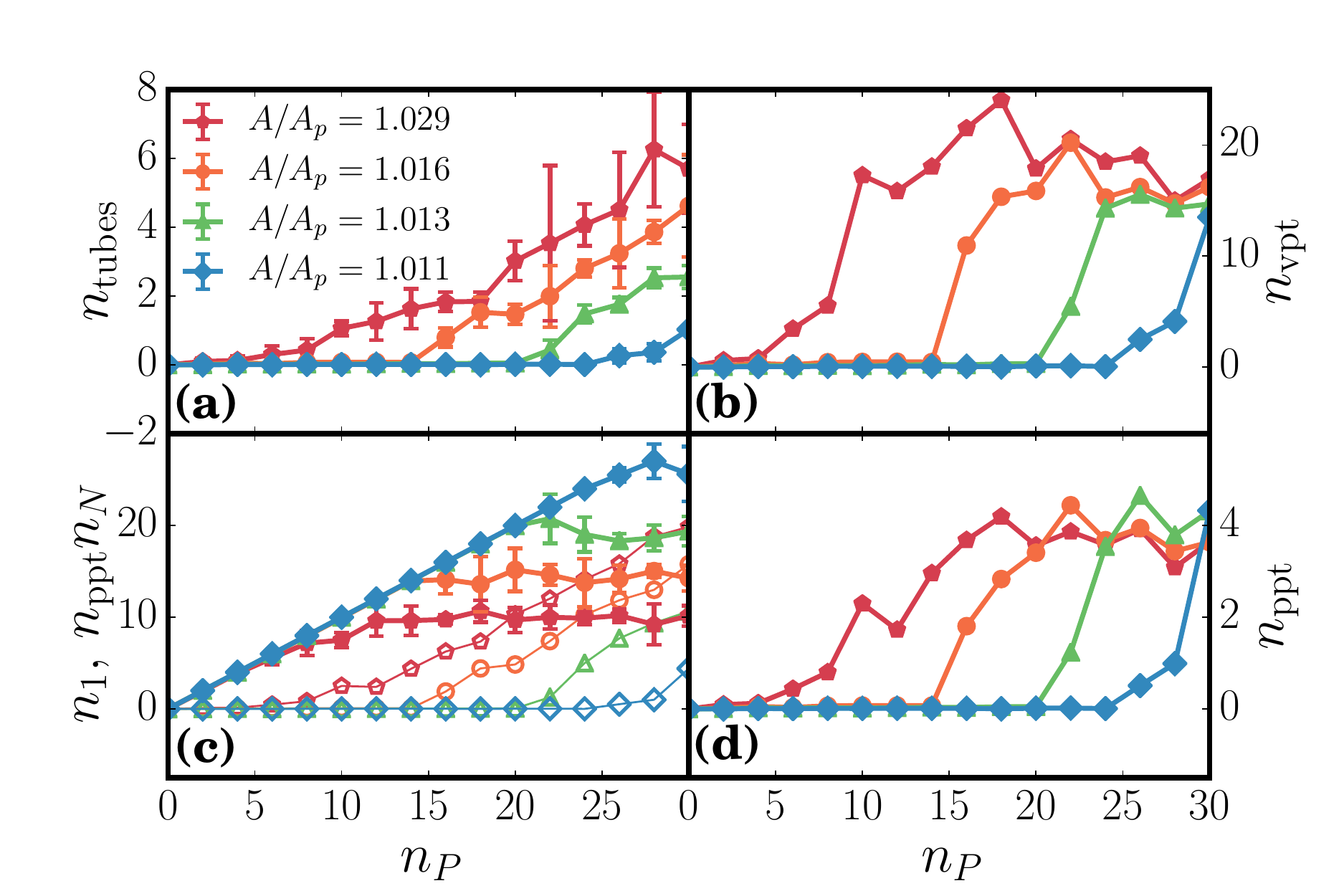}
\caption{(color online). Plot of several different tube statistics including a) the average number of tubes at each concentration for several excess areas $\left(n_{\rm tubes}\right)$, b) the average number of vertices per tubule $\left(n_{\rm vpt}\right)$, c) the average number of monomers $\left(n_1\right)$ and oligomers $\left(n_{\rm ppt} n_N\right)$ in simulation where monomers represent all proteins on the basal part of the membrane (closed symbols), and the n-mers represent all proteins in tubules (open symbols), and d) the average number of proteins per tubule $\left(n_{\rm ppt}\right)$. The legends in the panels correspond to four different values of $A/A_{p}$.
\label{fig:tube-statistics}}  
\end{figure}

The thermodynamics of tubule formation can be related to a critical aggregation concentration $n_{P,*}$, analogous to a critical micelle concentration (CMC).  An important parameter in micelle formation is the critical micelle number, or the number of surfactants in each micelle.  For tubule formation, this number is analogous to the number of membrane proteins in each tubule.  In our coarse-grained model for membranes, a single protein field represents $\zeta$ protein units and hence the  absolute number of proteins within each tubule is given by  $N_{\rm ppt}=n_{\rm ppt}\zeta$, where $n_{\rm ppt}$ is the number of coarse-grained protein fields in the tubular region. $n_{\rm ppt}$ as a function of the total number of coarse-grained proteins, $n_P$, for four different membrane excess areas, is shown in Fig.\ref{fig:tube-statistics}(d). It can be seen that $n_{\rm ppt}$ saturates to approximately 4, for all values of $n_{P}$ above a critical aggregation number $n_{P,*}$ whose value in turn depends on the elastic properties of the membrane and the parameters characterizing the protein field. 

In the classic analysis of micellar self-assembly \cite{Israelachvili:2011tr,Nels_2003_book} the total surfactant concentration ($c_{tot}$) is expressed in terms of the monomer concentration ($c_{1}$) and the concentration of an aggregate containing $M$ surfactant molecules ($c_{M}$) as,
\begin{eqnarray}
c_{tot} & = & c_1 + M c_M \\
& \equiv & c_1 \left( 1 + M c_1^{M-1} \left(\exp\left(M\beta(\mu_1^{0} - \mu_M^{0}) \right) \right) \right) \nonumber,
\label{eqn:israelachvilli-eqn}
\end{eqnarray}
with $(\mu_1^{0} - \mu_M^{0})$ being the chemical potential difference between the monomer state and the aggregate.  

In analogy, the proteins in the planar and tubular regions on the membrane correspond to the monomers and aggregates respectively. Thus following eqn.~\eqref{eqn:israelachvilli-eqn}, the equations governing the partitioning of proteins between the planar and tubular states can be rewritten in terms of the protein numbers as
\begin{equation}
\zeta n_{P}  =  \zeta n_1 +  \zeta n_{\rm ppt} n_N,
\label{eqn:tubule-micelle}
\end{equation}
with,
\begin{equation}
n_N=(\zeta n_1)^{\zeta n_{\rm ppt}} \left(\exp\left( \zeta n_{\rm ppt}\beta(\mu_p^{ex} - \mu_t^{ex}) \right) \right).
\label{eqn:expr-nn}
\end{equation}

$n_1$ is the number of protein fields in the planar phase (analogous to $c_{1}$), $n_N$ is the number of tubes each containing $\zeta n_{\rm ppt}$ proteins (analogous to the concentration of micelles $c_{M}$), and $\zeta n_{\rm ppt} n_N$ is the total number of proteins partitioned into the tubular phase. At the critical number of protein fields ($n_{P,*}$) that promotes membrane tubulation (see discussions by Nelson ~\cite{Nels_2003_book}),
\begin{equation}
n_{P} =  n_{P,*} \quad {\rm and} \quad n_{1}  =   n_{\rm ppt} n_{N}=n_{P,*}/2.
\label{eqn:cmc-def}
\end {equation}

Using eqns.~\eqref{eqn:expr-nn} and  ~\eqref{eqn:cmc-def}  in eqn.~\eqref{eqn:tubule-micelle} we obtain,
\begin{equation}
\zeta n_{\rm ppt} \exp \left(\beta \zeta n_{\rm ppt} \left (\mu_{p}^{ex}-\mu_{t}^{ex}\right ) \right)= \left (\dfrac{\zeta n_{P,*}}{2} \right )^{(1-\zeta n_{\rm ppt})}.
\label{eqn:cmc-constitutive}
\end{equation}

Thus, the number of protein fields in the planar and tubular regions are related through the equation,
\begin{equation}
n_{P} = n_1 \left( 1 + \left(\frac{2 n_1}{n_{P,*}}\right)^{N_{\rm ppt}-1} \right).
\label{eqn:tubule-micelle-2}
\end{equation}

\begin{figure}[!h] 
\centering
\includegraphics[width=7.5cm]{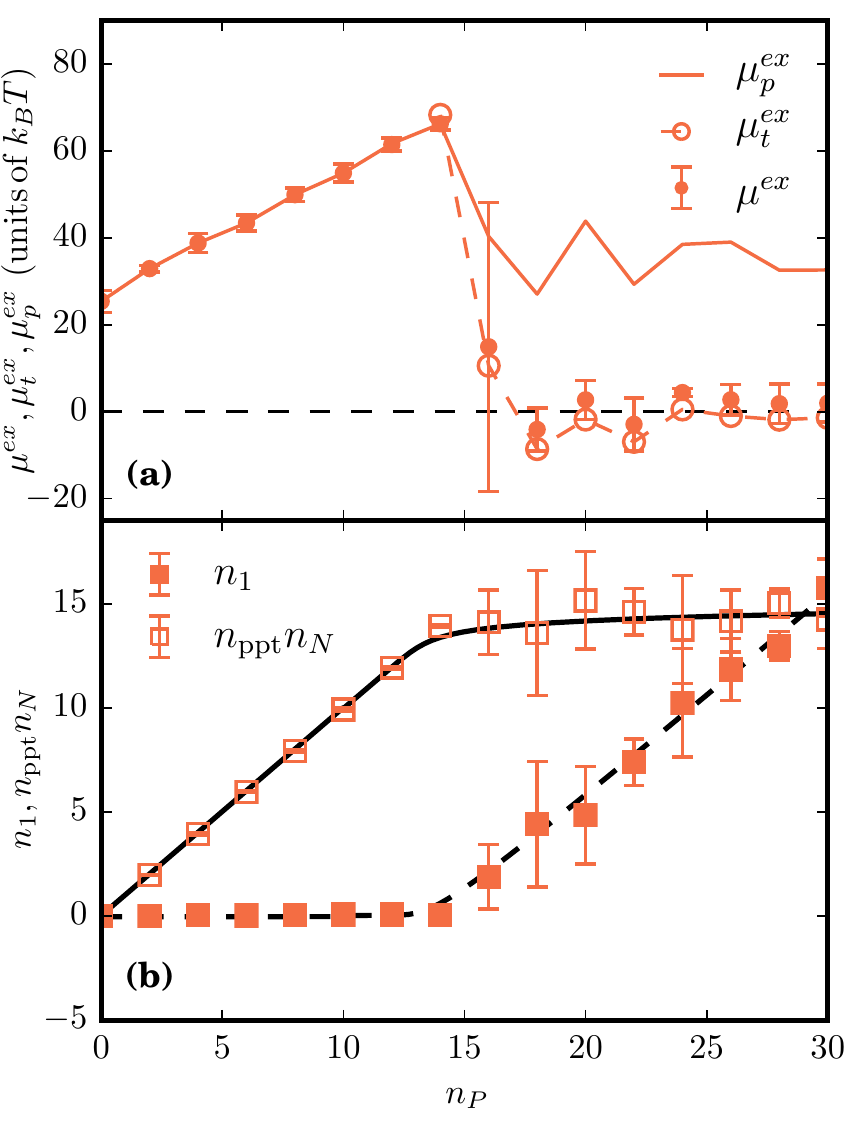}
\caption{(color online). (a) The various excess chemical potentials as a function of $n_P$, for $A/A_{p}$=1.016, $C_0=0.8$, and $\epsilon^2=6.3$. The filled symbols with error bars denote $\mu^{ex}$, open symbols with dotted lines represent $\mu_t^{ex}$, and solid lines correspond to $\mu_p^{ex}$. (b) Total number of protein fields in the planar ($n_1$) and  tubular ($n_{\rm ppt}n_N$) regions as a function of $n_P$. $n_{\rm ppt}$ corresponds to the average number of protein fields per tubule.  The solid and dashed black lines are the analytical fits to the micelle model described in eqn.~\eqref{eqn:tubule-micelle-2}.
\label{fig:micellization}}  
\end{figure}

Notice that despite being a coarse-grained model the number of coarse grained protein fields in the planar phase is related to the total number of proteins through the coarse graining parameter, $\zeta$, which appears in the exponent of eqn.~\eqref{eqn:tubule-micelle-2} on the right hand side. $\zeta$, as will be shown later, can be determined either by fitting the observed values of $n_{1}$ to eqn.~\eqref{eqn:tubule-micelle-2} or by analyzing how the critical protein density varies as a function of membrane tension, as shown in Fig.~\ref{fig:tensionvdensity} --- our scaling analysis yields a value for $\zeta=10$. Incidentally, this value of $\zeta$ shows an excellent fit of eqn.~\eqref{eqn:tubule-micelle-2} to our simulation data as shown in Fig.~\ref{fig:micellization}b. Methods to calculate the protein numbers in the planar and tubular regions are described below.

In order to compare the tubulation behavior in our simulations with eqn.~\eqref{eqn:tubule-micelle-2}, $n_{1}$, $n_{N}$ and $n_{\rm ppt}$ were calculated using a clustering algorithm with a mean curvature cutoff of $H = 0.5 \, a_0^{-1}$, similar to the cutoff used in inhomogeneous Widom insertion. The values of $n_{1}$, $n_{N}$ and $n_{\rm ppt}$, along with the number of vertices constituting a tube $n_{\rm vpt}$ are shown in Fig.~\ref{fig:tube-statistics}. All reported data are averaged over four independent ensembles each containing 150 uncorrelated membrane conformations.

The distinction between a phase transition in a finite system versus a state transition resulting in finite sized assemblies can be made by recognizing that the former would produce an ordered phase whose extent will span the size of the system. However, given that $\mu^{ex}$ in the tubular phase is flat with increasing $n_P$, following Israelachvili's argument~\cite{Israelachvili:2011tr}, multiple tubes of short (finite) lengths are entropically more favored rather than a single long tube, for which $\mu^{ex}$ versus $n_P$ should decrease monotonically post transition. The total number of proteins partitioned into the planar ($n_1$) and the tubular ($n_{\rm ppt}n_N$) regions, computed for a membrane with $A/A_p=1.016$, $C_0=0.8$, and $\epsilon^2=6.3$, are shown in Fig.~\ref{fig:micellization}; at the onset of tubulation, $n_1$ saturates and the number of proteins in the tubular regions increases linearly. A closer inspection of the tubule statistics (see Fig.~\ref{fig:tube-statistics}) reveals that with increasing $n_P$, the number of protein per tube remains fixed with $n_{\rm ppt}\approx 4$, while the number of tubes $n_N$ increases. These observations are characteristic of a micellization like transition and this is further evidenced in Fig.~\ref{fig:micellization} where our data shows excellent agreement with the predictions of the micellar model.  We rule out the possibility that the flat behavior of $\mu^{ex}$ versus $n_P$ is an artifact of our ensemble of holding $A_p$ fixed rather than maintaining a constant tension because the absolute value of the $\mu^{ex}$ of the tubular phase remains at a constant value for all values of $n_P$ post transition for systems with different $A_p$. Beyond providing insight into how the thermodynamic stability of the tubular phase is impacted by the independent variables $n_P$ and $A_p$, our results show that threshold density (the value of $n_P^{crit}$) that marks the onset of the tubular transition shifts to larger values with a decrease in the excess area $A/A_{p}$, which clearly implies that membrane tension $\sigma$ has a predominant effect on the transition. 

\subsection{Estimating membrane tension at tubulation}
\begin{figure}[!h] 
\centering
\includegraphics[width=7.5cm]{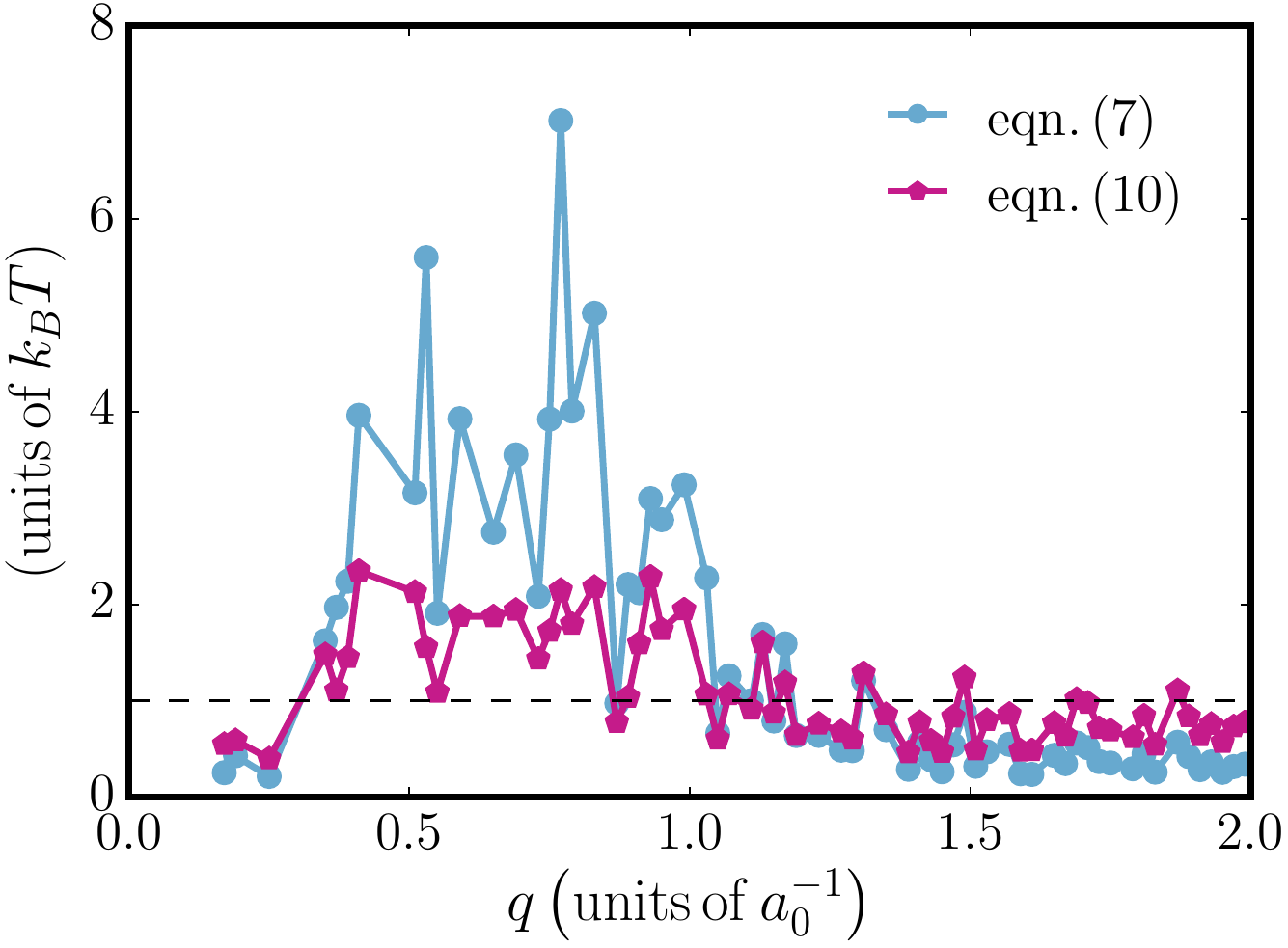}
\caption{(Color Online). Plots of the right hand sides eqn. \eqref{eqn:simple-hqhq} and eqn. \eqref{eqn:hqhmq-complex-3}, obtained by non-linear fitting procedures as a function of $q$. Data shown correspond to fits with a bin size of 0.02 and a maximum $q$ of 2, from a tubulated membrane corresponding to $\kappa=20$ \kbt,  $A/A_p = 1.029$, and  $n_P = 12$. 
\label{fig:simple-complex-fit}}  
\end{figure}

\begin{figure}[!h] 
\centering
\includegraphics[width=7.0cm]{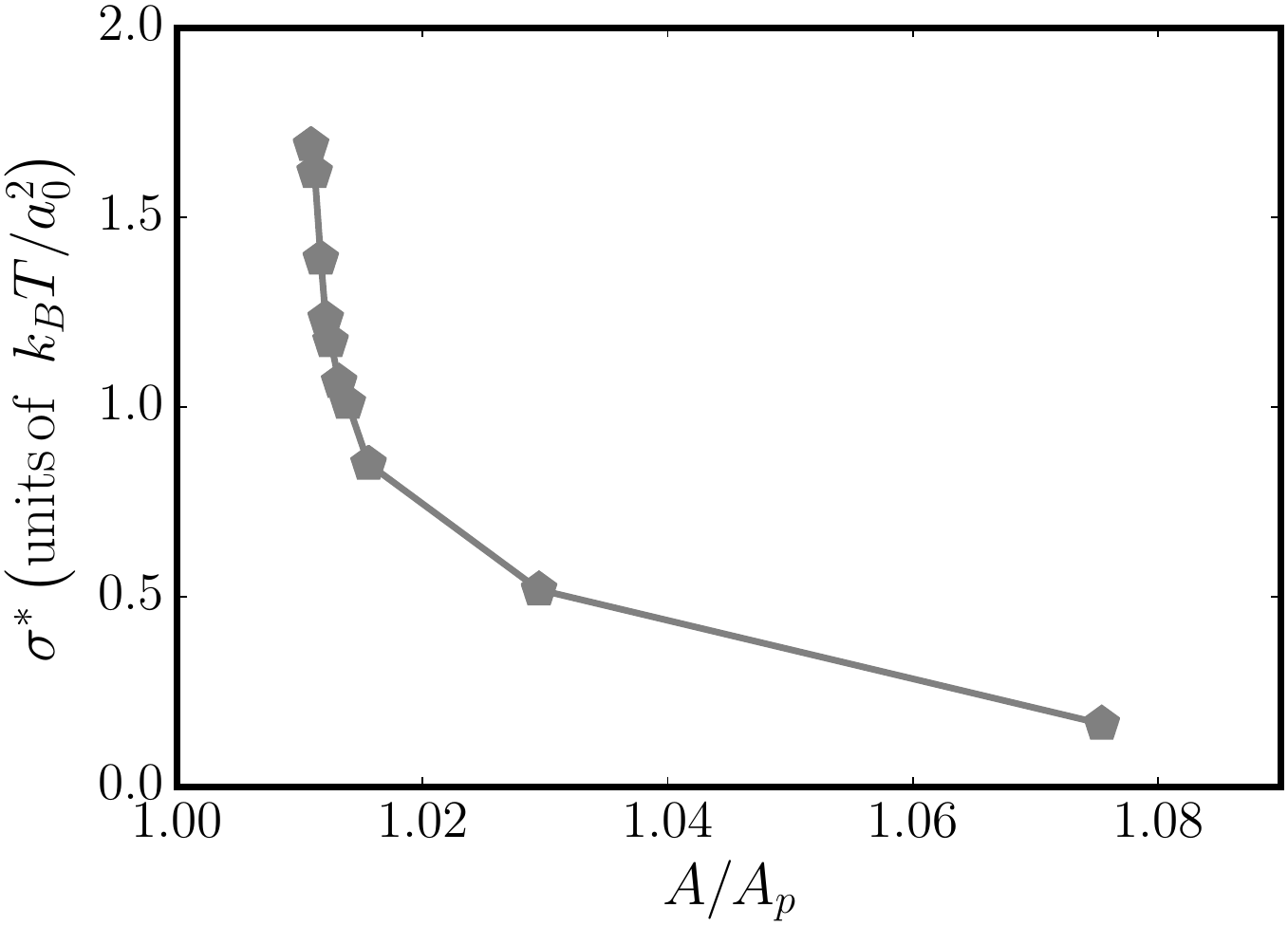}
\caption{(color online) Plot of $\sigma^{*}$, the membrane tension at tubulation as a function of $A/A_p$ for a membrane with $C_0=0.8 \, a_0^{-1}$ . \label{fig:sigma-at-tubulation}}  
\end{figure}

The membrane tension at the point of tubulation is an experimentally measurable quantity and the computational results can be compared to experiments if the tension at tubulation can be estimated accurately. As pointed out in Sec.~\ref{sec:uspec} the renormalized tension for planar membranes can be computed by analyzing their undulation spectrum. However, in the case of membranes with spontaneous curvature field, the long wavelength modes (i.e. small $q$) would violate equipartition if the conventional scaling relation given in eqn.~\eqref{eqn:simple-hqhq} is used. Hence, we  explicitly take the contributions from the spontaneous curvature field into account  and estimate $\sigma$ using eqn.~\eqref{eqn:hqhmq-complex-3}. A comparison of the equipartition relation for the best estimate of $\sigma$ determined using eqn.~\eqref{eqn:simple-hqhq} and eqn.~\eqref{eqn:hqhmq-complex-3} is shown in Fig.~\ref{fig:simple-complex-fit}, for a membrane with $\kappa=20$ \kbt{}, $A/A_p=1.029$ and $n_P=12$. It can be seen that the equipartition is better satisfied when the latter relation is used.  The values of $\sigma$, estimated using eqn.~\eqref{eqn:hqhmq-complex-3}, as a function of $n_P$ for various values of $A/A_p$ can be found in the Appendix~\ref{app:renorm-tension}. $\sigma^{*}$, the tension at tubulation is taken to be the value of membrane tension at the tubulation point, where the chemical potentials satisfy the condition $\mu_{p}^{ex}-\mu_{t}^{ex} \geq \mu^{ex}$. The membrane tension at the tubulation point as a function of $A/A_p$ for spontaneous curvature field with $C_0=0.8$ is shown in Fig.~\ref{fig:sigma-at-tubulation} and we observe that the tension for tubulation decreases with increasing excess area.

\subsection{Comparison of tension at tubulation to experiments}

\begin{figure}[h] 
\centering
\includegraphics[width=7.5cm]{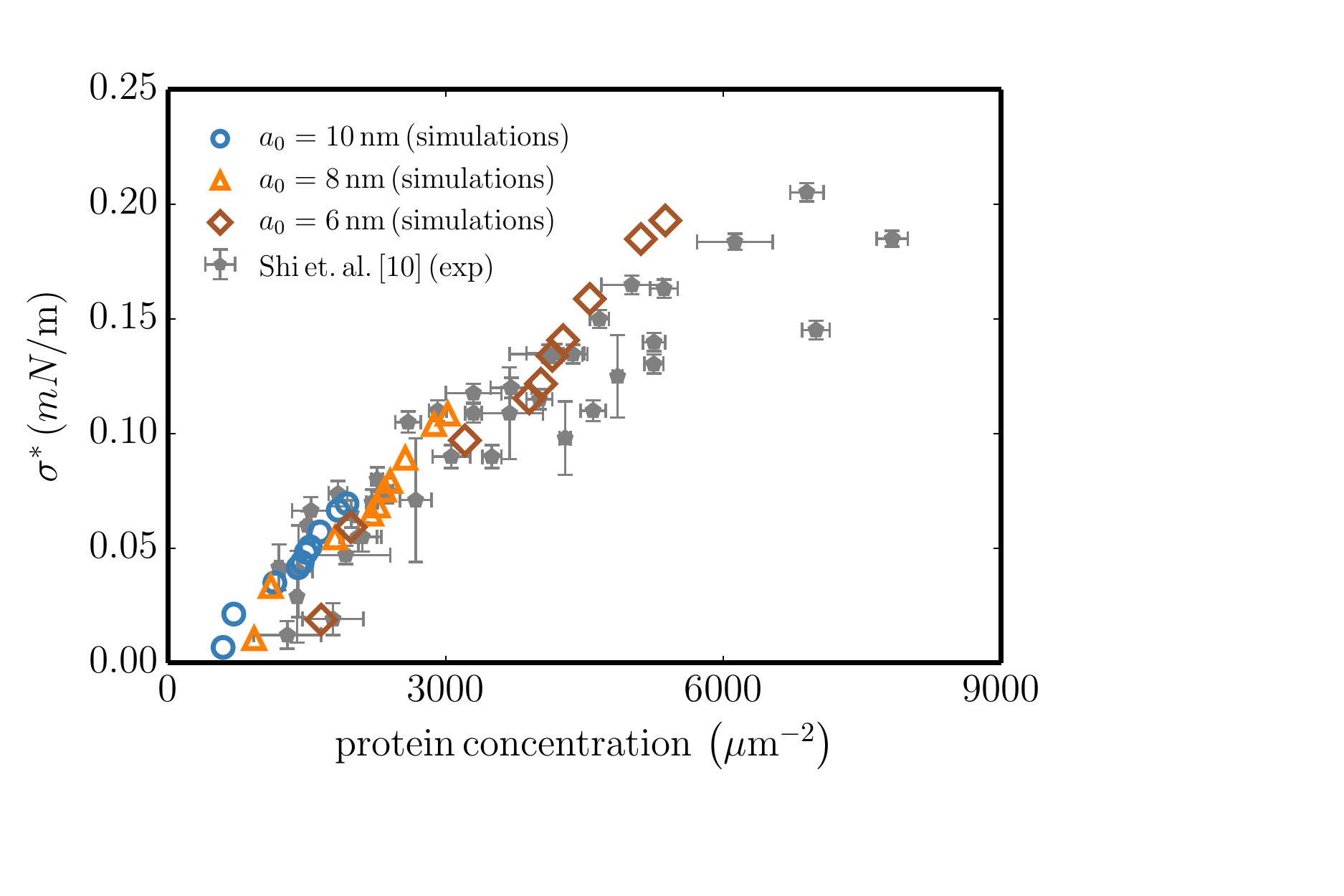}
\caption{(color online). Comparison of experimental (filled symbols)~\cite{Shi:2014ho} and simulation data (open 
symbols) for the averaged membrane tension and protein concentration at the point of tubulation. Simulation data are shown for three different values of the length scale $a_0$. In simulations, the protein concentration is  calculated as $\zeta n_P^{crit}/A_p$, where the coarse graining parameter $\zeta \approx 10$.
\label{fig:tensionvdensity}}  
\end{figure}

We test our model predictions against the critical tubulation density for endophilins reported 
by Shi and Baumgart ~\cite{Shi:2014ho}. Since, curvature-fields renormalize the values of 
$\sigma$, for a given $A$ the tension will depend on $n_P$ and differ from its value at $n_P=0$, we first develop a quantitative relationship between membrane area $A$ and membrane tension $\sigma$. In order to consider the effect of protein fields on renormalizing the tension values, we implement the modified fluctuation analysis method described in Sec.~\ref{sec:uspec}.  The computed values of the critical tension, $\sigma^{*}$ versus tubulation density are shown alongside the 
experimental data in Fig.~\ref{fig:tensionvdensity}. In order to make a direct comparison with experimental data, we 
self consistently determine the length scale $a_0$ by matching tubule diameters obtained in simulations to that in 
experiments~\cite{Ford:2002if,Peter:2004eb,Gallop:2006dm}, which yields values of $a_0$ in the range $6$ to $10\, {\rm 
nm}$. In turn, $a_0$ can be used to determine the corresponding protein density in our simulations, where each protein 
field is a coarse grained representation of $\zeta$ proteins, where $\zeta \geq 1$ can be regarded as the 
oligomerization number of protein domains needed to establish a stable curvature field. Estimated protein concentrations match those in experiments when the oligomerization parameter $\zeta \approx 10$ and we 
observe that the computed values of $\sigma^{*}$, for all values of $a_0$, are in good quantitative agreement with those measured from experiments. This estimate of $\zeta$ also matches extremely well with the value of the coarse grained parameter obtained through the micellar model, previously shown in Fig.~\ref{fig:micellization}(b).

In addition to $A/A_p$ (or membrane tension $\sigma$), both curvature field parameters $C_0$ and $\epsilon^2$ can also impact the onset of tubulation, as shown in Fig.~\ref{fig:mu-excess-parameters-bifurcation}, (see also Tables~\ref{table:czero} and~\ref{table:epsilonsq} in Appendix~\ref{app:delmu}). For weakly curving protein fields $C_0 < 0.6$, $\mu^{ex}$ shows a monotonic increase for the range $0<n_P<30$, implying the absence of a tubulation transition in this regime.  In contrast, when $C_0 > 0.6$, $\mu^{ex}$ displays the characteristic pitch-fork signature of tubulation, with the onset occurring at lower values of $n_P$ with for both $C_0 = 0.7$ and $0.8$. The critical tubulation density, however, remains unaltered with change in the value of $\epsilon^2$, see Fig.~\ref{fig:mu-excess-parameters-bifurcation}. Complementary to the critical tubulation density, ($n_{P}^{crit}$), we can estimate the saturation density of the proteins on the bilayer ($\rho^{max}$) using the relationship, $\rho^{max} \propto \exp{(-\mu^{max}/k_BT)}$~\cite{citeulike:4298529}, where $\mu^{max}$ is the value of the excess chemical potential just prior to tubulation; the values of $\mu^{max}$ for different $C_0, \epsilon^2, A/A_p$ are provided in Fig.~\ref{fig:mu-excess-parameters-bifurcation}, (see also Tables~\ref{table:czero} and~\ref{table:epsilonsq} in Appendix~\ref{app:delmu}). Based on our results, we find that $\rho^{max}$ and $n_{P}^{crit}$ both decrease with increasing $C_0$. Hence, proteins inducing a strong curvature field, can induce a morphological transition at lower densities, but also experience higher membrane-curvature mediated repulsive interactions, which limits their coverage on the membrane. The trends for $n_P^{crit}$ and $\rho^{max}$ versus $C_0$ as gleaned from our computed excess chemical potential landscape are currently being tested in experiments tracking membrane tubulation in three different protein systems. This predictive ability extends the utility of our model/simulations in defining the mechanisms of subtle yet important morphological transitions in soft biological systems, in delineating the 
thermodynamic stability of the underlying states; it further shows that the approach can be used to guide new experiments.  We advocate that this thermodynamic description at the microscopic 
resolution discussed here  will significantly impact and inform cellular mechanisms (including dynamics) mediated by 
emergent membrane morphologies driving intracellular trafficking and cell motility~\cite{Ramakrishnan:ureview}.

\acknowledgements{This work was supported in part by the National Science Foundation Grant DMR-1120901. The research leading to these results has received funding from the NIH grant 1U54CA193417. Computational resources were provided in part by the extreme science and engineering discovery environment (XSEDE) grant MCB060006.}

\appendix
\section{Renormalization of tension with protein number} \label{app:renorm-tension}
\begin{figure}[h] 
\centering
\includegraphics[width=7.5cm]{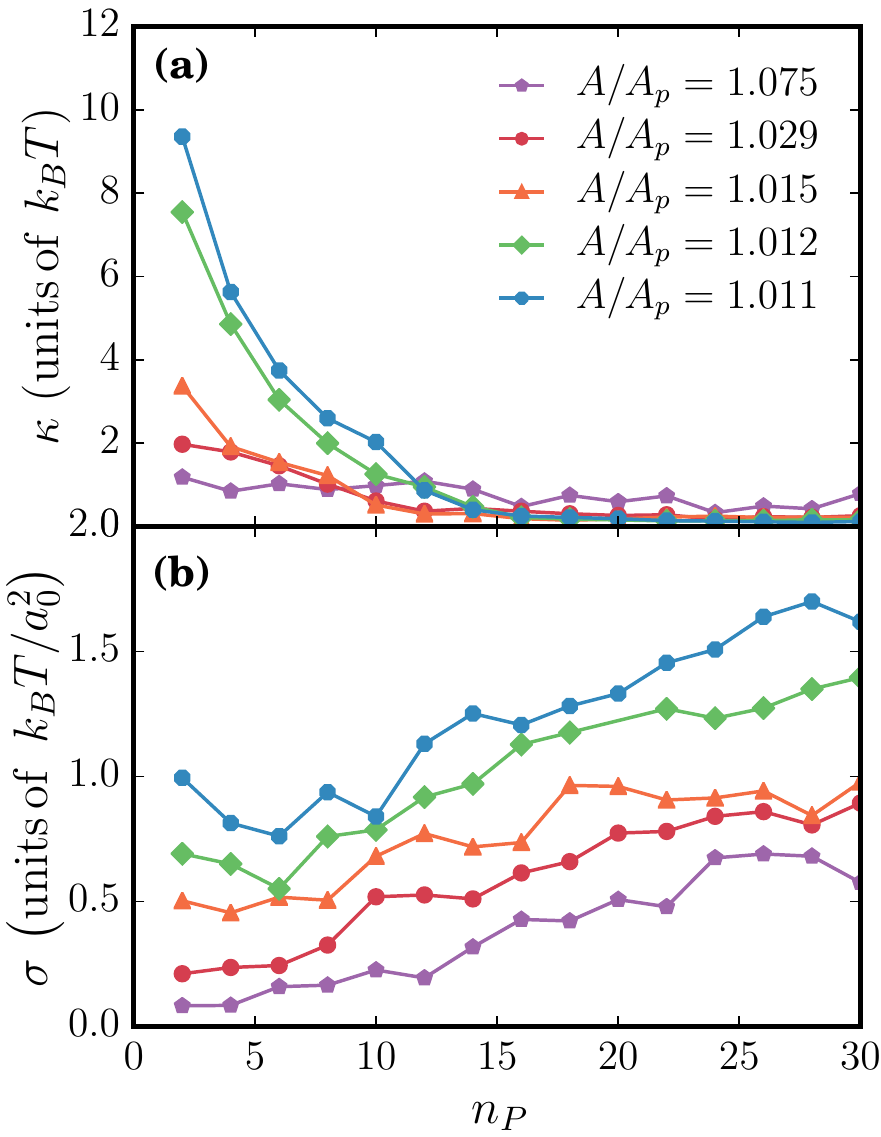}
\caption{(color online). Plot of the values of (a) $\kappa$ and (b) $\sigma$  obtained by nonlinear fitting of the complex spectrum eqn.~\eqref{eqn:hqhmq-complex-3} with tubules removed.  A bin size of 0.02 in $q$ and a maximum $q$ of 1 were used for these fits.
\label{fig:renormalized-values}}  
\end{figure}

As described before the renormalized values of $\kappa$ and $\sigma$, in the presence of spontaneous curvature inducing protein fields, can be determined through a nonlinear fit of eqn.\eqref{eqn:hqhmq-complex-3}.  Figs. \ref{fig:renormalized-values}a and \ref{fig:renormalized-values}b show the values of $\kappa$ and $\sigma$, estimated using eqn.\eqref{eqn:hqhmq-complex-3}, as a function of protein field number for several excess areas.  Since the Monge-Gauge approximation is valid only for small deformations we limit our analysis only to the planar regions on the membrane\textemdash in case of membranes with tubules these regions are neglected.  It can be seen in Fig.~\ref{fig:renormalized-values}b that the presence of proteins alters the in-plane undulatory modes of the membrane which is evidenced by an increase in the renormalized tension with increase in protein number.  As expected, the excess area and membrane tension are inversely related with the membrane sustaining high tension when the excess area reservoir is small and vice-versa, as shown in Fig.~\ref{fig:renormalized-values}b. Furthermore, we also observe that tensed membranes can be stabilized when the protein concentration is high and vice-versa.  On the other hand, our analysis shows that the membrane softens (i.e. $\kappa$ decreases) either with increase in excess area or protein concentration, which is shown in Fig.~\ref{fig:renormalized-values}a.  The value of tension at tubulation $(\sigma^*)$, defined as the tension of a membrane when $\mu_{p}^{ex}-\mu_{t}^{ex} \geq \mu^{ex}$, points to the fact that the membrane requires a critical excess area for tubulation transitions to occur.  This can be seen in Fig.~\ref{fig:sigma-at-tubulation} which shows the divergence of $\sigma^{*}$ at smaller values of $A/A_{p}$.  

\section{ $\langle \mu_p^{ex} - \mu_t^{ex} \rangle$ dependence curvature field parameters } \label{app:delmu}
The critical density for tubulation shows a dependence on both membrane tension, and the curvature field parameters $C_0$ and $\epsilon^2$.  Plots of the various chemical potentials, $\mu^{ex}$, $\mu_p^{ex}$, and $\mu_t^{ex}$, as a function of $C_0$, $\epsilon^2$ and $A/A_{p}$ are shown in  Fig.~\ref{fig:mu-excess-parameters-bifurcation}.  The critical number of protein fields required to stabilize membrane regions with mean curvatures above the cutoff value of $H>0.5a_{0}^{-1}$ is a strong function of $C_{0}$ and $\epsilon^{2}$. It should be noted that depending on the value of $C_{0}$, the regions corresponding to $H>0.5a_{0}^{-1}$ can either be blebs (a spherical bud) or tubules, with the former being predominant for $C_{0} \approx 0.6a_{0}^{-1}$ and the latter being stable for $C_{0} \geq 0.8a_{0}^{-1}$ (see \cite{cite:moviessi}). The formation of regions with curvatures above the cutoff is accompanied by a drop in the value of chemical potential $\mu^{ex}$ as seen in all the panels in Fig.~\ref{fig:mu-excess-parameters-bifurcation}. The scaling of $\mu^{ex}$ preceding tubulation is consistent with earlier results reported in Tourdot et. al.~\cite{Tourdot:2014ef}.

The excess chemical potential $\mu^{ex}$ increases with increase in $n_{P}$ and peaks at $n_{P}=n_{P}^{crit}$, with peak value $\mu_{max}$. The critical number of protein fields required to form blebs or tubes is taken to be the value of $n_{P}=n_{P}^{crit}$ at which this drop occurs. However, the values of $n_{P}^{crit}$ can be also determined by analyzing the behavior of the various chemical potentials. We take $n_{P}^{crit}$ to be the minimum value of $n_{P}$ at which the chemical potentials obey the relation $\mu^{ex}_{p}-\mu^{ex}_{t}>\mu^{ex}$. Tables.~\ref{table:czero} and ~\ref{table:epsilonsq} show the values of the various chemical potentials and critical protein number for various systems shown in Fig.~\ref{fig:mu-excess-parameters-bifurcation}.

The Widom insertion technique gives reliable estimates for the chemical potentials for a wide range of parameters characterizing the membrane-protein system especially when the mean curvature distributions, $P(H)$, show a broad distribution whose range is much greater than $C_{0}/2$. It should be noted that when a protein field with spontaneous curvature $C_{0}$ is inserted on a membrane surface the dominant contributions to $\mu^{ex}$ come from membrane regions with $2H \approx C_{0}$.

Hence, in analyzing the effects of $C_{0}$ and $\epsilon^{2}$ on the morphological transitions, we only consider values of  $A/A_{p}>1.013$, which clearly satisfy this criterion for $P(H)$, see Tables.~\ref{table:czero} and ~\ref{table:epsilonsq}, for our results.

\begin{table*}[!h]
\caption{\label{table:czero} Values of $\mu^{max}$, $\mu^{ex}_{p}-\mu^{ex}_{t}$, and $n_{p}^{crit}$ as a function of $C_{0}$ and $A/A_{p}$ for fixed value of $\epsilon^{2}=6.3a_{0}^{2}$.  Values of (-) represent parameters where no tubules were observed or less than three values were obtained to in order calculate the corresponding standard deviation.}
\centering
\begin{tabular}{ccccc}
\hline
 & \quad $C_{0}$ \quad &  \quad $\langle \mu^{ex}_{p}-\mu^{ex}_{t} \rangle_{n_P > n_P^{crit}}$ \quad & \quad $\mu^{max}$ \quad & \quad $\, n_{P}^{crit} \,$ \quad\\
$\, A/A_{p} \,$ & (units of $a_{0}^{-1}$) & (units of $k_{B}T$) & (units of $k_{B}T$) & $\, ( \pm 1) \,$ \\
\hline 
\multirow{4}{4em}{\centering 1.029} & 0.5 & 11.7 $\pm$ 3.0 & 9.8 $\pm$ 6.6 & 14 \\ 
& 0.6 & 17.2 $\pm$ 4.8 & 16.0 $\pm$ 5.6 & 15 \\ 
& 0.7 & 24.5 $\pm$ 3.9 & 19.5 $\pm$ 7.5 & 5 \\
& 0.8 & 28.5 $\pm$ 3.2 & 41.7 $\pm$ 3.9 & 6 \\ 
\hline
\multirow{4}{4em}{\centering 1.016} & 0.5 & 14.1 $\pm$ 3.1 & 26.4 $\pm$ 1.5 & 22 \\ 
& 0.6 & 23.2 $\pm$ 3.1 & 33.5 $\pm$ 6.3 & 16 \\ 
& 0.7 & 24.2 $\pm$ 4.3 & 34.8 $\pm$ 2.2 & 15 \\
& 0.8 & 29.3 $\pm$ 3.6 & 72.8 $\pm$ 3.9 & 15 \\ 
\hline
\multirow{4}{4em}{\centering 1.013} & 0.5 & - & - & - \\ 
& 0.6 & 28.9 $\pm$ - & 46.1 $\pm$ 6.7 & 24 \\ 
& 0.7 & 25.0 $\pm$ 6.0 & 44.3 $\pm$ 2.0 & 18 \\
& 0.8 & 51.4 $\pm$ 3.8 & 80.4 $\pm$ 1.2 & 22 \\ 
\hline
\end{tabular}
\end{table*}

\begin{table*}[!h]
\caption{\label{table:epsilonsq} Values of $\mu^{max}$, $\mu^{ex}_{p}-\mu^{ex}_{t}$, and $n_{p}^{crit}$ as a function of $\epsilon^{2}$  and $A/A_{p}$ for fixed value of $C_{0}=0.8a_{0}^{-1}$. Values of (-) represent parameters where no tubules were observed or less than three values were obtained in order to calculate the corresponding standard deviation.}
\centering
\begin{tabular}{ccccc} 
\hline
& \quad $\epsilon^{2}$ \quad & \quad $\langle \mu^{ex}_{p}-\mu^{ex}_{t} \rangle_{n_P > n_P^{crit}}$ \quad &\quad  $\mu^{max}$ \quad & \quad $\, n_{P}^{crit} \,$ \quad \\
$\, A/A_{p} \,$ & (units of $a_{0}^{2}$) & (units of $k_{B}T$) & (units of $k_{B}T$) & $\, ( \pm 1)�\,$ \\
\hline 
\multirow{4}{4em}{\centering 1.029} & 2.3 & 9.4 $\pm$ 1.8 & 4.6 $\pm$ 1.8 & 8 \\ 
& 4.3 & 23.4 $\pm$ 3.0 & 11.7 $\pm$ 7.1 & 5 \\ 
& 6.3 & 30.6 $\pm$ 4.1 & 46.4 $\pm$ 4.1 & 8 \\
& 8.3 & 33.2 $\pm$ 3.2 & 73.5 $\pm$ 8.2 & 12 \\ 
\hline
\multirow{4}{4em}{\centering 1.016} & 2.3 & 12.1 $\pm$ 3.9 & 10.4 $\pm$ 0.8 & 16 \\ 
& 4.3 & 28.2 $\pm$ 5.3 & 29.3 $\pm$ 1.0 & 12 \\ 
& 6.3 & 42.8 $\pm$ 15.1 & 62.1 $\pm$ 1.9 & 16 \\
& 8.3 & 48.8 $\pm$ 11.7 & 107.6 $\pm$ 7.4 & 14 \\ 
\hline
\multirow{4}{4em}{\centering 1.013} & 2.3 & 13.6 $\pm$ - & 15.5 $\pm$ 0.4 & 28 \\ 
& 4.3 & 36.2 $\pm$ 4.2 & 36.8 $\pm$ 1.4 & 18 \\ 
& 6.3 & 48.9 $\pm$ 8.1 & 79.4 $\pm$ 2.7 & 18 \\
& 8.3 & 60.3 $\pm$ 13.6 & 134.4 $\pm$ 0.7 & 20 \\ 
\hline
\bottomrule
\end{tabular}
\end{table*}

%

\end{document}